\documentclass[%
 onecolumn,
 notitlepage,
 longbibliography,
 amsmath,amssymb,
 aps,
 pra,
 10pt,
]{revtex4-1}

\usepackage{graphicx}
\usepackage{dcolumn}
\usepackage{bm}
\usepackage{bbold}
\usepackage{braket}
\usepackage{mathtools}

\begin{document}

\title{Derivation of the quantum-optical master equation based on coarse-graining of time}

\author{Kevin Fischer}
 \affiliation{E. L. Ginzton Laboratory, Stanford University, Stanford CA 94305, USA}
 \email{kevinf@stanford.edu}

\date{\today}

\begin{abstract}
This is a derivation of the quantum-optical master equation using coarse-graining of time, which brings new insights into a decades old technique. My derivation is quite similar to derivations using quantum stochastic methods or Kraus operators, though I go through the derivation without explicitly invoking any of these concepts, so it may be easier to follow as an introduction. I also address the major pitfall of nearly all microscopic derivations of the master equation, namely that they assume the state of the system and bath factorize for all times. I show why this assumption actually holds for spontaneous emission, and coincidentally turns out to be correct.
\end{abstract}

\maketitle

\section{Schr\"{o}dinger-picture Hamiltonian}

Consider a chiral (unidirectional) waveguide with a single spatial mode profile, supporting a bath of harmonic oscillators. In the quasi-monochromatic regime, where the frequency content of the excitations in the waveguide is narrowband \cite{fischer2017scattering}, its Hamiltonian is approximately
\begin{eqnarray}
H_0 = \mathbb{1}_\text{sys}\otimes\int_{-\infty}^{\infty} \mathop{\textrm{d}\omega}\omega b_\omega^\dagger b_\omega .
\end{eqnarray}
If the waveguide is linearly coupled to a local quantum system, with the Hamiltonian $H_\text{sys}$, then in the rotating wave approximation the interaction part of the Hamiltonian is
\begin{eqnarray}
H_1 = H_\text{sys}\otimes\mathbb{1}_\text{field} + \textrm{i}\sqrt{\gamma}\left(\sigma \otimes \int_{-\infty}^{\infty} \frac{\mathop{\textrm{d}\omega}}{\sqrt{2\pi}}b_\omega^\dagger -\sigma^\dagger \otimes \int_{-\infty}^{\infty} \frac{\mathop{\textrm{d}\omega}}{\sqrt{2\pi}}b_\omega \right).
\end{eqnarray}
The field mode operators obey the commutation relation
\begin{eqnarray}
[b_\omega, b^\dagger_{\omega'}]=\delta(\omega-\omega'),\label{eq:commute0}
\end{eqnarray}
and $\sigma$ is a lowering operator in the Hilbert space of the system. This gives a total Hamiltonian
\begin{eqnarray}
H=H_0+H_1.
\end{eqnarray}
The quantum-optical master equation describes the reduced dynamics of the system alone after having traced out the field modes
\begin{equation}
\rho_\text{sys}(t)=\text{Tr}_\text{field}\big[\ket{\Psi(t)}\bra{\Psi(t)}\big] \quad\text{ or }\quad \rho_\text{sys}(t)=\text{Tr}_\text{field}\big[\rho(t)\big],
\end{equation}
depending on the initial conditions. It has been widely used in quantum optics and is responsible for many of the field's successes. Further, it has been derived in many forms, though in my opinion they all have numerous shortcomings:
\begin{enumerate}
\item The original microscopic derivation \cite{carmichael2009open} has the seemingly unphysical assumption that the state of the bath and system factorize at all times, i.e. $\rho(t)=\rho_\text{sys}(t)\otimes\rho_\text{field}(t)$. This is obviously not true since it is easily possible to maximally entangle the system and field, e.g. in a spin-photon interface \cite{gao2012observation,yao2005theory}.
\item The measurement-theory derivation usually invokes a similarly dubious assumption \cite{jacobs2014quantum,wiseman2009quantum}, though recently that was shown unnecessary \cite{gross2017qubit}.
\item Methods based on stochastic evolution equations require significant knowledge of stochastic calculus which most quantum mechanics students do not possess \cite{gardiner2004quantum}.
\item Information-theoretic derivations based on the properties of dynamical semigroups \cite{breuer2002theory,haroche2006exploring} do not immediately reveal how physical models can be cast into the same form as their results.
\end{enumerate}
In this tutorial, I aim to address these issues by borrowing techniques and concepts from measurement theory, while basing the calculations exclusively on a microscopic model. I hope this presentation will be helpful for those trying to understand the quantum-optical master equation.

\section{interaction-picture Hamiltonian}

Due to the coupling between the local system and all frequency modes of the waveguide, integrating Schr\"{o}dinger's equation for this Hamiltonian is a bit unwieldy. The first step to getting a Hamiltonian that is easier to work with is to transform into an interaction picture to remove the free evolution of the waveguide $H_0$. Specifically, I choose the state vector
\begin{equation}
\ket{\Psi_\textrm{I}(t)}=\textrm{e}^{\textrm{i}H_{0}t}\ket{\Psi(t)}.
\end{equation}
Then, Schr\"{o}dinger's equation for $\ket{\Psi_\textrm{I}(t)}$ becomes
\begin{equation}
\textrm{i}\frac{\partial}{\partial t}\ket{\Psi_\textrm{I}(t)}=H_\textrm{I}(t)\ket{\Psi_\textrm{I}(t)}
\end{equation}
with
\begin{eqnarray}
H_\textrm{I}(t)&=&\textrm{e}^{\textrm{\scriptsize i}H_{0}t}  H_1\textrm{e}^{-\textrm{\scriptsize i}H_{0} t}\\
&=&H_\text{sys}\otimes\mathbb{1}_\text{field} + \textrm{i}\sqrt{\gamma}\left(\sigma \otimes \int_{-\infty}^{\infty} \frac{\mathop{\textrm{d}\omega}}{\sqrt{2\pi}}\textrm{e}^{\textrm{i}\omega t} b_\omega^\dagger -\sigma^\dagger \otimes \int_{-\infty}^{\infty} \frac{\mathop{\textrm{d}\omega}}{\sqrt{2\pi}}\textrm{e}^{-\textrm{i}\omega t} b_\omega \right).
\end{eqnarray}
Here, I used the relationship
\begin{eqnarray}
\textrm{e}^{\textrm{i}H_{0}t} b_\omega \textrm{e}^{-\textrm{i}H_{0}t}&=&\textrm{e}^{-\textrm{i}\omega t}b_\omega,
\end{eqnarray}
which follows from the commutation in Eq. \ref{eq:commute0}. Next, I define a new operator, which naturally appeared as a Fourier transform of the $b_\omega$'s
\begin{eqnarray}
b(t) &=& \int_{-\infty}^{\infty} \frac{\mathop{\textrm{d}\omega}}{\sqrt{2\pi}}\textrm{e}^{-\textrm{i}\omega t} b_\omega.
\end{eqnarray}
I then rewrite the interaction-picture Hamiltonian as
\begin{eqnarray}
H_\textrm{I}(t)&=&H_\text{sys}\otimes\mathbb{1}_\text{field} + V(t),\label{eq:HI_t}
\end{eqnarray}
where
\begin{eqnarray}
V(t)=\textrm{i}\sqrt{\gamma}\left(\sigma \otimes b^\dagger(t) -\sigma^\dagger \otimes b(t) \right).\label{eq:Vt}
\end{eqnarray}
The operator $b(t)$ obeys commutations similar to the frequency mode operators, but in time
\begin{eqnarray}
[b(t), b^\dagger(t')]&=&\left[\int_{-\infty}^{\infty} \frac{\mathop{\textrm{d}\omega}}{\sqrt{2\pi}}\textrm{e}^{-\textrm{i}\omega t} b_\omega,\int_{-\infty}^{\infty} \frac{\mathop{\textrm{d}\omega'}}{\sqrt{2\pi}}\textrm{e}^{\textrm{i}\omega' t'} b_{\omega'}^\dagger\right]\\
&=&\int_{-\infty}^{\infty} \int_{-\infty}^{\infty}\frac{\mathop{\textrm{d}\omega}}{\sqrt{2\pi}} \frac{\mathop{\textrm{d}\omega'}}{\sqrt{2\pi}}\textrm{e}^{-\textrm{i}\omega t}\textrm{e}^{\textrm{i}\omega' t'}\left[ b_\omega, b_{\omega'}^\dagger\right]\\
&=&\int_{-\infty}^{\infty} \int_{-\infty}^{\infty}\frac{\mathop{\textrm{d}\omega}}{\sqrt{2\pi}} \frac{\mathop{\textrm{d}\omega'}}{\sqrt{2\pi}}\textrm{e}^{-\textrm{i}\omega t}\textrm{e}^{\textrm{i}\omega' t'}\delta(\omega-\omega')\\
&=&\int_{-\infty}^{\infty} \frac{\mathop{\textrm{d}\omega}}{2\pi} \textrm{e}^{-\textrm{i}\omega (t-t')}\\
&=&\delta(t-t').
\end{eqnarray}

\section{coarse-grained time evolution operator}

The Schr\"{o}dinger equation can be rewritten by iteratively applying its integral form, yielding a Dyson series
\begin{eqnarray}
\ket{\Psi_\textrm{I}(t_1)}&=&\ket{\Psi_\textrm{I}(t_0)}+\int_{t_0}^{t_1}\mathop{\text{d}t}\frac{\partial}{\partial t}\ket{\Psi_\textrm{I}(t)}\\
&=&\ket{\Psi_\textrm{I}(t_0)}-\textrm{i}\int_{t_0}^{t_1}\mathop{\text{d}t}H_\textrm{I}(t)\ket{\Psi_\textrm{I}(t)}\\
&=&\ket{\Psi_\textrm{I}(t_0)}-\textrm{i}\int_{t_0}^{t_1}\mathop{\text{d}t}H_\textrm{I}(t)\ket{\Psi_\textrm{I}(t_0)}+(-\text{i})^2\int_{t_0}^{t_1}\mathop{\text{d}t}\int_{t_0}^{t}\mathop{\text{d}t'}H_\textrm{I}(t)H_\textrm{I}(t')\ket{\Psi_\textrm{I}(t)}\\
&=&\left(\mathbb{1}_\text{sys}\otimes \mathbb{1}_\text{field}-\textrm{i}\int_{t_0}^{t_1}\mathop{\text{d}t}H_\textrm{I}(t)+(-\text{i})^2\int_{t_0}^{t_1}\mathop{\text{d}t}\int_{t_0}^{t}\mathop{\text{d}t'}H_\textrm{I}(t)H_\textrm{I}(t')+\cdots\right)\ket{\Psi_\textrm{I}(t_0)}
\end{eqnarray}
Note that the limits on the second time index $t'$ are only from $t_0<t'<t$ and not $t_0<t'<t_1$. Since the arguments of each set of integrals are symmetric with respect to exchange of time indices, I can alternatively let $t'$ vary until $t_1$ by applying the time-ordering operator $\mathcal{T}$. On the second-order term it is defined by
\begin{eqnarray}
\mathcal{T}H_\textrm{I}(t)H_\textrm{I}(t')=\begin{cases}
H_\textrm{I}(t)H_\textrm{I}(t')&\text{ if } t>t'\\
H_\textrm{I}(t')H_\textrm{I}(t)&\text{ if } t<t'
\end{cases},
\end{eqnarray}
yielding the restatement
\begin{eqnarray}
\ket{\Psi_\textrm{I}(t_1)}&=&\left(\mathbb{1}_\text{sys}\otimes \mathbb{1}_\text{field}-\textrm{i}\int_{t_0}^{t_1}\mathop{\text{d}t}H_\textrm{I}(t)+\frac{(-\text{i})^2}{2!}\int_{t_0}^{t_1}\mathop{\text{d}t}\int_{t_0}^{t_1}\mathop{\text{d}t'}\mathcal{T}H_\textrm{I}(t)H_\textrm{I}(t')+\cdots\right)\ket{\Psi_\textrm{I}(t_0)}\label{eq:24}\\
&\equiv&\mathcal{T}\text{exp}\left[-\text{i}\int_{t_0}^{t_1}\mathop{\text{d}t} H_\textrm{I}(t)\right]\ket{\Psi_\textrm{I}(t_0)}.
\end{eqnarray}
This unitary operator $U(t_1,t_0)=\mathcal{T}\text{exp}\left[-\text{i}\int_{t_0}^{t_1}\mathop{\text{d}t} H_\textrm{I}(t)\right]$ is called a time-ordered exponential and is a shorthand for the expansion in Eq. \ref{eq:24}.

Now, I can proceed to finding an approximate dynamical map that takes the wavefunction from time $t=k\Delta t\rightarrow (k+1)\Delta t$, i.e. the map $U[k+1,k]$ where
\begin{equation}
\ket{\Psi_\textrm{I}[k+1]} = U[k+1,k] \ket{\Psi_\textrm{I}[k]}.
\end{equation}
The goal of this map is to remove the time-ordering (which is intractable for this Hamiltonian), while maintaining unitarity (so the norm of the wavefuctnion is conserved) and being correct to $\mathcal{O}(\Delta t)$ (so that the error vanishes in the limit of $\Delta t\rightarrow 0$) \cite{fischer2017scattering}. Hence, I choose
\begin{equation}
U[k+1,k] \approx \textrm{exp}\left[-\textrm{i}\int_{k\Delta t}^{(k+1)\Delta t}\mathop{\textrm{d}t} H_\textrm{I}(t)\right].
\end{equation}
This map differs from the correct map via time-ordering
\begin{eqnarray}
\textrm{Error}&=&U[k+1,k]-U((k+1)\Delta t, k\Delta t)\\
&=&\textrm{exp}\left[-\textrm{i}\int_{k\Delta t}^{(k+1)\Delta t}\mathop{\textrm{d}t} H_\textrm{I}(t)\right] - \mathcal{T} \textrm{exp}\left[-\textrm{i}\int_{k\Delta t}^{(k+1)\Delta t}\mathop{\textrm{d}t} H_\textrm{I}(t)\right]\\
&=&\int_{k\Delta t}^{(k+1)\Delta t}\mathop{\textrm{d}t}\int_{t}^{(k+1)\Delta t}\mathop{\textrm{d}t'}\left[H_\textrm{I}(t),H_\textrm{I}(t')\right] + \cdots,\label{eq:erroranalysis}
\end{eqnarray}
where the limits of integration are only over the upper half of the coordinate plane for $t<t'$. Here, the operators $H_\textrm{I}(t)$ and $H_\textrm{I}(t')$ need to be reordered, which gives their commutator in Eq. \ref{eq:erroranalysis}. I can break up the commutation between the Hamiltonian at different times based on Eq. \ref{eq:HI_t}
\begin{eqnarray}
\left[H_\textrm{I}(t),H_\textrm{I}(t')\right]&=&\left[H_\textrm{sys}(t),H_\textrm{sys}(t')\right]+\left[H_\textrm{sys}(t),V(t')\right]+\left[V(t),H_\textrm{sys}(t')\right]+\left[V(t),V(t')\right]\\
&=& \qquad\quad\,\mathcal{O}(1)\quad\quad\;+\;\;\;\;\mathcal{O}(1/\sqrt{\Delta t})\;\,+\;\;\mathcal{O}(1/\sqrt{\Delta t})\;\;\;\:+\quad \quad 0.
\end{eqnarray}
The commutations between $H_\textrm{sys}(t)$ and $V(t')$ are assigned an $\mathcal{O}(1/\sqrt{\Delta t})$ because the singular operators in $V(t)$ (Eq. \ref{eq:Vt}) obey the commutation
\begin{equation}
\left[\int_{k\Delta t}^{(k+1)\Delta t}\mathop{\textrm{d}t} b(t),\int_{k\Delta t}^{(k+1)\Delta t}\mathop{\textrm{d}t'} b^\dagger(t')\right]=\Delta t.
\end{equation}
Combining the orders of $\left[H_\textrm{I}(t),H_\textrm{I}(t')\right]$ in Eq. \ref{eq:erroranalysis}, I see that my choice of map is correct to $\mathcal{O}(\Delta t)$, with a leading error $\mathcal{O}(\Delta t^{3/2})$ from the commutation of $H_\textrm{sys}(t)$ and $V(t')$. This approximation amounts to a coarse-graining of the system-bath-interaction dynamics to a timescale of $\Delta t$, which occur on a much faster timescale than the dynamics generated by the system evolution. Nevertheless, in the limit of $\Delta t\rightarrow 0$ the error vanishes and the map becomes exact.

Writing the coarse-grained map out explicitly
\begin{eqnarray}
U[k+1,k]&=&\textrm{exp}\left[-\textrm{i}H_\text{sys}\Delta t\otimes\mathbb{1}_\text{field}+\sqrt{\gamma\Delta t}\Big(\sigma \otimes\int_{k\Delta t}^{(k+1)\Delta t} \mathop{\textrm{d}t} \frac{b^\dagger(t)}{\sqrt{\Delta t}} - \sigma^\dagger \otimes\int_{k\Delta t}^{(k+1)\Delta t} \mathop{\textrm{d}t}\frac{b(t)}{\sqrt{\Delta t}}\Big)\right]\\
&=&\textrm{exp}\left[-\textrm{i}H_\text{sys}\Delta t\otimes\mathbb{1}_\text{field}+\sqrt{\gamma\Delta t}\Big(\sigma \otimes\Delta B^\dagger[k] - \sigma^\dagger\otimes\Delta B[k] \Big)\right],
\end{eqnarray}
where I defined the coarse-grained operator
\begin{equation}
\Delta B[k]=\int_{k\Delta t}^{(k+1)\Delta t}\mathop{\textrm{d}t} \frac{b(t)}{\sqrt{\Delta t}}.
\end{equation}
This operator obeys the commutation
\begin{equation}
\left[\Delta B[j],\Delta B^\dagger[k]\right]=\delta_{jk}\label{eq:commute}
\end{equation}
and in the limit
\begin{equation}
\lim_{\Delta t\rightarrow 0}\frac{\Delta B[t/\Delta t]}{\sqrt{\Delta t}} = b(t).
\end{equation}
(A very nice alternative to this derivation is based on using a wavelet expansion of $b(t)$, which yields an approximate Hamiltonian directly rather than just a coarse-grained map \cite{gross2017qubit}.)

Now, the accessible Hilbert space of the field is also coarse-grained. Specifically, the map can only create or remove excitations from the field in time bins of width $\Delta t$. Owing to the commutation in Eq. \ref{eq:commute}, the relevant Hilbert space of the field is now a product of a bunch of harmonic oscillator spaces, each labeled by the time-bin number $n$ 
\begin{equation}
\mathcal{H}_\textrm{field}^\textrm{coarse}=\bigotimes_{n=-\infty}^{+\infty} \mathcal{H}_\infty^n.
\end{equation}
The vacuum state with zero excitations is $\ket{\mathbf{0}}=\ket{0}\otimes\ket{0}\otimes\cdots$, the field with a single excitation is in a state
\begin{eqnarray}
\Delta B^\dagger[j]\ket{\mathbf{0}} &=& \cdots\otimes\ket{ 0}\otimes\ket{ 0}\otimes\ket{1_j}\otimes\ket{ 0}\otimes\ket{ 0}\otimes\cdots,
\end{eqnarray}
or with multiple excitations in the same $j$-th bin
\begin{eqnarray}
\frac{\left(\Delta B^\dagger[j]\right)^m}{\sqrt{m!}}\ket{\mathbf{0}} &=& \cdots\otimes\ket{ 0}\otimes\ket{ 0}\otimes\ket{m_j}\otimes\ket{ 0}\otimes\ket{ 0}\otimes\cdots.
\end{eqnarray}

\section{quantum-optical master equation for spontaneous emission}

Consider the initial state of the local system to be a mixed state, with the field in the vacuum state
\begin{eqnarray}
\rho(t_0=0)&=&\rho_\text{sys}(0)\otimes\ket{\mathbf{0}}\bra{\mathbf{0}}.
\end{eqnarray}
To obtain the density matrix of the system at time $t_1=\Delta t$, I need to evolve the total system with the unitary map $U[1,0]$ and then trace out the field states
\begin{eqnarray}
\rho_\text{sys}[1]&=&\text{Tr}_\text{field}\Big[U[1,0]\{\rho_\text{sys}[0]\otimes\ket{\mathbf{0}}\bra{\mathbf{0}}\}U[0,1]\Big]\label{eq:43}
\end{eqnarray}
The first step to evaluating this expression is to expand the trace as a series of partial traces over the field time bins, with the partial trace over the $j$-th bin as
\begin{eqnarray}
\text{tr}_\text{bin j}\big[\cdots\big]&\equiv&\sum_m \bra{0_j}\frac{\left(\Delta B[j]\right)^m}{\sqrt{m!}} \cdots\frac{\left(\Delta B^\dagger[j]\right)^m}{\sqrt{m!}}\ket{0_j},
\end{eqnarray}
so
\begin{eqnarray}
\rho_\text{sys}[1]&=&\text{tr}_\text{bin 0}\text{tr}_\text{bin 1}\cdots\text{tr}_{\text{bin }\infty}\Big[U[1,0]\{\rho_\text{sys}[0]\otimes\ket{\mathbf{0}}\bra{\mathbf{0}}\}U[0,1]\Big].\label{eq:4444}
\end{eqnarray}
Then, I note that note that $U[k+1,k]$ only acts on $\mathcal{H}_\textrm{sys}\otimes\mathcal{H}_{n=k}$ so
\begin{equation}
\left[\Delta B[j],U[k,j+1]\right]=0\quad\textrm{and}\quad
\left[\Delta B[j],U[j,l]\right]=0.\label{eq:55}
\end{equation}
Thus, the creation and annihilation operators from partial traces over bins other than the $0$-th one in Eq. \ref{eq:4444} commute with the unitary evolution operators and evaluate to zero in their expectations. This leaves only the first partial trace and the zero elements from the other partial traces---hence
\begin{eqnarray}
\rho_\text{sys}[1]&=&\cdots\otimes\bra{0_2}\otimes\bra{0_1}\text{tr}_\text{bin 0}\Big[U[1,0]\{\rho_\text{sys}[0]\otimes\ket{\mathbf{0}}\bra{\mathbf{0}}\}U[0,1]\Big]\ket{0_1}\otimes\ket{0_2}\otimes\cdots\\
&=&\sum_m \bra{\mathbf{0}}\frac{\left(\Delta B[0]\right)^m}{\sqrt{m!}} U[1,0]\ket{\mathbf{0}}\rho_\text{sys}[0]\bra{\mathbf{0}}U[0,1]\frac{\left(\Delta B^\dagger[0]\right)^m}{\sqrt{m!}}\ket{\mathbf{0}}\label{eq:46}.
\end{eqnarray}
Notably this map is of the operator-sum representation \cite{jacobs2014quantum}, i.e.
\begin{eqnarray}
\rho_\text{sys}[1]&=&\sum_m K_{m}[0]\rho_\text{sys}[0]K_{m}^\dagger[0],\label{eq:47}
\end{eqnarray}
where $K_{m}[k]$ are the so-called Kraus operators (which are typically associated with quantum measurements, though I do not invoke any measurement theory in my derivation).

In my specific scenario, the Kraus operators for the first time step $\Delta t$ in evolution are
\begin{eqnarray}
K_{m}[0]&=&\bra{\mathbf{0}}\frac{\left(\Delta B[0]\right)^m}{\sqrt{m!}} U[1,0]\ket{\mathbf{0}}.
\end{eqnarray}
To evaluate these operators, I expand the discrete map to $\mathcal{O}(\Delta t)$
\begin{eqnarray}
U[1,0]&=&\textrm{exp}\left[-\textrm{i}H_\text{sys}\Delta t\otimes\mathbb{1}_\text{field}+\sqrt{\gamma\Delta t}\Big(\sigma \otimes\Delta B^\dagger[0] - \sigma^\dagger\otimes\Delta B[0] \Big)\right]\\
&=&\mathbb{1}_\text{sys}\otimes \mathbb{1}_\text{field} + \sqrt{\gamma\Delta t}\Big(\sigma \otimes\Delta B^\dagger[0] - \sigma^\dagger\otimes\Delta B[0] \Big)+ \nonumber\\
&&\hspace{80pt} \Delta t\Big[-\text{i}H_\text{sys}\otimes\mathbb{1}_\text{field} +\frac{1}{2}\gamma\Big(\sigma \otimes\Delta B^\dagger[0] - \sigma^\dagger\otimes\Delta B[0] \Big)^2\Big] + \mathcal{O}(\Delta t^{3/2}).
\end{eqnarray}
Then, I evaluate these Kraus operators for zero photon emissions into the field
\begin{eqnarray}
K_{0}[0]&=&\bra{\mathbf{0}}U[1,0]\ket{\mathbf{0}}\\
&= & \mathbb{1}_\text{sys}+\Delta t\left(-\text{i}H_\text{sys} -\frac{\gamma}{2}\sigma^\dagger\sigma\right) + \mathcal{O}(\Delta t^{3/2}),
\end{eqnarray}
one photon emission into the field
\begin{eqnarray}
K_{1}[0]&=&\bra{\mathbf{0}}\Delta B[0]U[1,0]\ket{\mathbf{0}}\\
&= & \sqrt{\gamma\Delta t}\, \sigma + \mathcal{O}(\Delta t^{3/2}),
\end{eqnarray}
and two photon emissions into the field
\begin{eqnarray}
K_{2}[0]&=&\bra{\mathbf{0}}\frac{(\Delta B[0])^2}{\sqrt{2}}U[1,0]\ket{\mathbf{0}}\\
&= & \mathcal{O}(\Delta t).
\end{eqnarray}
Operators representing more photon emissions are higher order in $\Delta t$ and hence are taken $K_{m>2}[0]\approx 0$. Then, keeping all terms $\mathcal{O}(\Delta t)$ in the density matrix map
\begin{eqnarray}
\rho_\text{sys}[1]&=&\sum_m K_{m}[0]\rho_\text{sys}[0]K_{m}^\dagger[0]\\
&\approx&\rho_\text{sys}[0]+\Delta t\left[\left(-\text{i}H_\text{sys} -\frac{\gamma}{2}\sigma^\dagger\sigma\right)\rho_\text{sys}[0]+\rho_\text{sys}[0]\left(\text{i}H_\text{sys}-\frac{\gamma}{2}\sigma^\dagger\sigma\right) + \gamma \sigma\rho_\text{sys}[0]\sigma^\dagger\right].
\end{eqnarray}
Rearranging, I show this is in the standard Lindblad form
\begin{eqnarray}
\frac{\rho_\text{sys}[1]-\rho_\text{sys}[0]}{\Delta t}&=&-\text{i}\left[H_\text{sys}, \rho_\text{sys}[0]\right] + \gamma\left(\sigma \rho_\text{sys}[0]\sigma^\dagger-\frac{1}{2}\{\sigma^\dagger\sigma,\rho_\text{sys}[0]\}\right).
\end{eqnarray}

It is further important to show that this map holds for all time steps. For example, consider the second time step, i.e. from $t=0\rightarrow\Delta t\rightarrow 2\Delta t$.
\begin{eqnarray}
\rho_\text{sys}[2]&=&\text{Tr}_\text{field}\Big[U[2,0]\{\rho_\text{sys}[0]\otimes\ket{\mathbf{0}}\bra{\mathbf{0}}\}U[0,2]\Big]\\
&=&\text{tr}_\text{bin 0}\text{tr}_\text{bin 1}\cdots\text{tr}_{\text{bin }\infty}\Big[U[2,0]\{\rho_\text{sys}[0]\otimes\ket{\mathbf{0}}\bra{\mathbf{0}}\}U[0,2]\Big]\\
&=&\cdots\otimes\bra{0_3}\otimes\bra{0_2}\text{tr}_\text{bin 1}\text{tr}_\text{bin 0}\Big[U[2,0]\{\rho_\text{sys}[0]\otimes\ket{\mathbf{0}}\bra{\mathbf{0}}\}U[0,2]\Big]\ket{0_2}\otimes\ket{0_3}\otimes\cdots\\
&=&\sum_{m'}\sum_m \bra{\mathbf{0}}\frac{\left(\Delta B[1]\right)^{m'}}{\sqrt{m'!}} \frac{\left(\Delta B[0]\right)^m}{\sqrt{m!}}U[2,0]\ket{\mathbf{0}}\rho_\text{sys}[0]\bra{\mathbf{0}}U[0,2]\frac{\left(\Delta B^\dagger[1]\right)^{m'}}{\sqrt{m'!}}\frac{\left(\Delta B^\dagger[0]\right)^m}{\sqrt{m!}}\ket{\mathbf{0}}\label{eq:64}
\end{eqnarray}
Again, I have used Eq. \ref{eq:55} to reduce the complexity of the trace by commuting away operators that have no effect on the expectations. Next, I chronologically order and group all operators based on their action on the time bins and insert a field identity operator $\mathbb{1}_\text{field}$ between the groups. Then, I note that all the elements of the identity with nonzero photon number can commute past the other operators, either to the left or right depending on their time bin, and annihilate to zero. Hence, I can equivalently insert $\ket{\mathbf{0}}\bra{\mathbf{0}}$ rather than the identity. The left expectation in Eq. \ref{eq:64} then becomes
\begin{eqnarray}
\bra{\mathbf{0}}\frac{\left(\Delta B[1]\right)^{m'}}{\sqrt{m'!}} \frac{\left(\Delta B[0]\right)^m}{\sqrt{m!}}U[2,0]\ket{\mathbf{0}}&=&\bra{\mathbf{0}}\Big\{\frac{\left(\Delta B[1]\right)^{m'}}{\sqrt{m'!}}U[2,1]\Big\}\mathbb{1}_\text{field}\Big\{\frac{\left(\Delta B[0]\right)^m}{\sqrt{m!}}U[1,0]\Big\}\ket{\mathbf{0}}\\
&=&\bra{\mathbf{0}}\frac{\left(\Delta B[1]\right)^{m'}}{\sqrt{m'!}}U[2,1]\ket{\mathbf{0}}\bra{\mathbf{0}}\frac{\left(\Delta B[0]\right)^m}{\sqrt{m!}}U[1,0]\ket{\mathbf{0}}.
\end{eqnarray}

Applying to all of Eq. \ref{eq:64}, I obtain
\begin{eqnarray}
\rho_\text{sys}[2]&=&\sum_{m'}\sum_m \bra{\mathbf{0}}\frac{\left(\Delta B[1]\right)^{m'}}{\sqrt{m'!}} U[2,1]\ket{\mathbf{0}}\bra{\mathbf{0}}\frac{\left(\Delta B[0]\right)^m}{\sqrt{m!}} U[1,0]\ket{\mathbf{0}}\rho_\text{sys}[0]\bra{\mathbf{0}}U[0,1]\frac{\left(\Delta B^\dagger[0]\right)^m}{\sqrt{k!}}\ket{\mathbf{0}}\bra{\mathbf{0}}U[1,2]\frac{\left(\Delta B^\dagger[1]\right)^{m'}}{\sqrt{m'!}}\ket{\mathbf{0}}\nonumber\\
&=&\text{Tr}_\text{field}\Big[U[2,1]\{\rho_\text{sys}[1]\otimes\ket{\mathbf{0}}\bra{\mathbf{0}}\}U[1,2]\Big]\label{eq:65},
\end{eqnarray}
by inspection of Eq. \ref{eq:46}. Note that Eq. \ref{eq:65} looks just like Eq. \ref{eq:43}, but with all indices stepped by one and hence all times stepped by $\Delta t$! This pattern (or recursion relation) then holds for all future time steps as well, based on the ability to insert $\ket{\mathbf{0}}\bra{\mathbf{0}}$ between operators acting on different time bins. Another way to state this result is that the Kraus operators are time-independent, i.e. $K_{m}[0]=K_{m}[k]\equiv K_m$, and hence the quantum system has a type of Markovian evolution. Physically, the intuition is that the local system interacts with each temporal bin only once and thus the state of the bath modes after interaction are irrelevant if we consider only the reduced dynamics of the system. Here, we are free to choose the interacted modes to be vacuum again and get the same reduced dynamics as if we kept track of their entire state. \textit{This justifies the off-hand remark in nearly all quantum optics texts that} $\rho(t)=\rho_\text{sys}(t)\otimes\rho_\text{field}(t)$, \textit{which otherwise is not mathematically supported}. Hence,
\begin{eqnarray}
\frac{\rho_\text{sys}[k+1]-\rho_\text{sys}[k]}{\Delta t}&=&-\text{i}\left[H_\text{sys}, \rho_\text{sys}[k]\right] + \gamma\left(\sigma \rho_\text{sys}[k]\sigma^\dagger-\frac{1}{2}\{\sigma^\dagger\sigma,\rho_\text{sys}[k]\}\right)
\end{eqnarray}
and in the continuum limit
\begin{eqnarray}
\frac{\partial}{\partial t}\rho_\text{sys}(t)&=&-\text{i}\left[H_\text{sys}, \rho_\text{sys}(t)\right] + \gamma\left(\sigma \rho_\text{sys}(t)\sigma^\dagger-\frac{1}{2}\{\sigma^\dagger\sigma,\rho_\text{sys}(t)\}\right).\\
\end{eqnarray}
This is often written with the Liouvillian superoperator $\mathcal{L}$
\begin{eqnarray}
\frac{\partial}{\partial t}\rho_\text{sys}(t)&=&\mathcal{L}\rho_\text{sys}(t)\\
&=&-\text{i}\left[H_\text{sys}, \rho_\text{sys}(t)\right] + \mathcal{D}[\sqrt{\gamma}\sigma]\rho_\text{sys}(t)
\end{eqnarray}
and the so-called Dissipator
\begin{eqnarray}
 \mathcal{D}[\sqrt{\gamma}\sigma]\rho_\text{sys}(t)=\gamma\left(\sigma \rho_\text{sys}(t)\sigma^\dagger-\frac{1}{2}\{\sigma^\dagger\sigma,\rho_\text{sys}(t)\}\right).
\end{eqnarray}
Notably, the entire derivation holds for other system operators like $\sigma^\dagger\sigma$ to yield a dephasing rate instead of spontaneous emission. See reference \cite{jacobs2014quantum} for a detailed explanation of how thermal bath states drive the system, and their corresponding Lindblad superoperators.

In summary, I derived the quantum-optical master equation from a microscopic model of linear system-bath interactions, without relying on the unsupported assumption that the system and bath factorize at all times. I hope the techniques presented here will be useful in helping students enter the field of open quantum systems.

\section{Acknowledgements}

I gratefully acknowledge financial support from the National Science Foundation (Division of Materials Research---Grant. No. 1503759), and for feedback and discussions with Jelena Vu\v{c}kovi\'c.


\bibliography{bibliography}

\begin{thebibliography}{10}%
\makeatletter
\providecommand \@ifxundefined [1]{%
 \@ifx{#1\undefined}
}%
\providecommand \@ifnum [1]{%
 \ifnum #1\expandafter \@firstoftwo
 \else \expandafter \@secondoftwo
 \fi
}%
\providecommand \@ifx [1]{%
 \ifx #1\expandafter \@firstoftwo
 \else \expandafter \@secondoftwo
 \fi
}%
\providecommand \natexlab [1]{#1}%
\providecommand \enquote  [1]{``#1''}%
\providecommand \bibnamefont  [1]{#1}%
\providecommand \bibfnamefont [1]{#1}%
\providecommand \citenamefont [1]{#1}%
\providecommand \href@noop [0]{\@secondoftwo}%
\providecommand \href [0]{\begingroup \@sanitize@url \@href}%
\providecommand \@href[1]{\@@startlink{#1}\@@href}%
\providecommand \@@href[1]{\endgroup#1\@@endlink}%
\providecommand \@sanitize@url [0]{\catcode `\\12\catcode `\$12\catcode
  `\&12\catcode `\#12\catcode `\^12\catcode `\_12\catcode `\%12\relax}%
\providecommand \@@startlink[1]{}%
\providecommand \@@endlink[0]{}%
\providecommand \url  [0]{\begingroup\@sanitize@url \@url }%
\providecommand \@url [1]{\endgroup\@href {#1}{\urlprefix }}%
\providecommand \urlprefix  [0]{URL }%
\providecommand \Eprint [0]{\href }%
\providecommand \doibase [0]{http://dx.doi.org/}%
\providecommand \selectlanguage [0]{\@gobble}%
\providecommand \bibinfo  [0]{\@secondoftwo}%
\providecommand \bibfield  [0]{\@secondoftwo}%
\providecommand \translation [1]{[#1]}%
\providecommand \BibitemOpen [0]{}%
\providecommand \bibitemStop [0]{}%
\providecommand \bibitemNoStop [0]{.\EOS\space}%
\providecommand \EOS [0]{\spacefactor3000\relax}%
\providecommand \BibitemShut  [1]{\csname bibitem#1\endcsname}%
\let\auto@bib@innerbib\@empty
\bibitem [{\citenamefont {Fischer}\ \emph {et~al.}(2017)\citenamefont
  {Fischer}, \citenamefont {Trivedi}, \citenamefont {Ramasesh}, \citenamefont
  {Siddiqi},\ and\ \citenamefont {Vu{\v{c}}kovi{\'c}}}]{fischer2017scattering}%
  \BibitemOpen
  \bibfield  {author} {\bibinfo {author} {\bibfnamefont {Kevin~A}\ \bibnamefont
  {Fischer}}, \bibinfo {author} {\bibfnamefont {Rahul}\ \bibnamefont
  {Trivedi}}, \bibinfo {author} {\bibfnamefont {Vinay}\ \bibnamefont
  {Ramasesh}}, \bibinfo {author} {\bibfnamefont {Irfan}\ \bibnamefont
  {Siddiqi}}, \ and\ \bibinfo {author} {\bibfnamefont {Jelena}\ \bibnamefont
  {Vu{\v{c}}kovi{\'c}}},\ }\bibfield  {title} {\enquote {\bibinfo {title}
  {Scattering of coherent pulses from quantum-optical systems},}\ }\href@noop
  {} {\bibfield  {journal} {\bibinfo  {journal} {arXiv preprint
  arXiv:1710.02875}\ } (\bibinfo {year} {2017})}\BibitemShut {NoStop}%
\bibitem [{\citenamefont {Carmichael}(2009)}]{carmichael2009open}%
  \BibitemOpen
  \bibfield  {author} {\bibinfo {author} {\bibfnamefont {Howard}\ \bibnamefont
  {Carmichael}},\ }\href@noop {} {\emph {\bibinfo {title} {An open systems
  approach to quantum optics: lectures presented at the Universit{\'e} Libre de
  Bruxelles, October 28 to November 4, 1991}}},\ Vol.~\bibinfo {volume} {18}\
  (\bibinfo  {publisher} {Springer Science \& Business Media},\ \bibinfo {year}
  {2009})\BibitemShut {NoStop}%
\bibitem [{\citenamefont {Gao}\ \emph {et~al.}(2012)\citenamefont {Gao},
  \citenamefont {Fallahi}, \citenamefont {Togan}, \citenamefont
  {Miguel-S{\'a}nchez},\ and\ \citenamefont {Imamoglu}}]{gao2012observation}%
  \BibitemOpen
  \bibfield  {author} {\bibinfo {author} {\bibfnamefont {WB}~\bibnamefont
  {Gao}}, \bibinfo {author} {\bibfnamefont {Parisa}\ \bibnamefont {Fallahi}},
  \bibinfo {author} {\bibfnamefont {Emre}\ \bibnamefont {Togan}}, \bibinfo
  {author} {\bibfnamefont {Javier}\ \bibnamefont {Miguel-S{\'a}nchez}}, \ and\
  \bibinfo {author} {\bibfnamefont {Atac}\ \bibnamefont {Imamoglu}},\
  }\bibfield  {title} {\enquote {\bibinfo {title} {Observation of entanglement
  between a quantum dot spin and a single photon},}\ }\href@noop {} {\bibfield
  {journal} {\bibinfo  {journal} {Nature}\ }\textbf {\bibinfo {volume} {491}},\
  \bibinfo {pages} {426} (\bibinfo {year} {2012})}\BibitemShut {NoStop}%
\bibitem [{\citenamefont {Yao}\ \emph {et~al.}(2005)\citenamefont {Yao},
  \citenamefont {Liu},\ and\ \citenamefont {Sham}}]{yao2005theory}%
  \BibitemOpen
  \bibfield  {author} {\bibinfo {author} {\bibfnamefont {Wang}\ \bibnamefont
  {Yao}}, \bibinfo {author} {\bibfnamefont {Ren-Bao}\ \bibnamefont {Liu}}, \
  and\ \bibinfo {author} {\bibfnamefont {LJ}~\bibnamefont {Sham}},\ }\bibfield
  {title} {\enquote {\bibinfo {title} {Theory of control of the spin-photon
  interface for quantum networks},}\ }\href@noop {} {\bibfield  {journal}
  {\bibinfo  {journal} {Physical review letters}\ }\textbf {\bibinfo {volume}
  {95}},\ \bibinfo {pages} {030504} (\bibinfo {year} {2005})}\BibitemShut
  {NoStop}%
\bibitem [{\citenamefont {Jacobs}(2014)}]{jacobs2014quantum}%
  \BibitemOpen
  \bibfield  {author} {\bibinfo {author} {\bibfnamefont {Kurt}\ \bibnamefont
  {Jacobs}},\ }\href@noop {} {\emph {\bibinfo {title} {Quantum measurement
  theory and its applications}}}\ (\bibinfo  {publisher} {Cambridge University
  Press},\ \bibinfo {year} {2014})\BibitemShut {NoStop}%
\bibitem [{\citenamefont {Wiseman}\ and\ \citenamefont
  {Milburn}(2009)}]{wiseman2009quantum}%
  \BibitemOpen
  \bibfield  {author} {\bibinfo {author} {\bibfnamefont {Howard~M}\
  \bibnamefont {Wiseman}}\ and\ \bibinfo {author} {\bibfnamefont {Gerard~J}\
  \bibnamefont {Milburn}},\ }\href@noop {} {\emph {\bibinfo {title} {Quantum
  measurement and control}}}\ (\bibinfo  {publisher} {Cambridge university
  press},\ \bibinfo {year} {2009})\BibitemShut {NoStop}%
\bibitem [{\citenamefont {Gross}\ \emph {et~al.}(2018)\citenamefont {Gross},
  \citenamefont {Caves}, \citenamefont {Milburn},\ and\ \citenamefont
  {Combes}}]{gross2017qubit}%
  \BibitemOpen
  \bibfield  {author} {\bibinfo {author} {\bibfnamefont {Jonathan~Arthur}\
  \bibnamefont {Gross}}, \bibinfo {author} {\bibfnamefont {Carlton~M}\
  \bibnamefont {Caves}}, \bibinfo {author} {\bibfnamefont {Gerard~J}\
  \bibnamefont {Milburn}}, \ and\ \bibinfo {author} {\bibfnamefont {Joshua}\
  \bibnamefont {Combes}},\ }\bibfield  {title} {\enquote {\bibinfo {title}
  {Qubit models of weak continuous measurements:: Markovian conditional and
  open-system dynamics},}\ }\href@noop {} {\bibfield  {journal} {\bibinfo
  {journal} {Quantum Science and Technology}\ }\textbf {\bibinfo {volume}
  {3}},\ \bibinfo {pages} {024005} (\bibinfo {year} {2018})}\BibitemShut
  {NoStop}%
\bibitem [{\citenamefont {Gardiner}\ and\ \citenamefont
  {Zoller}(2004)}]{gardiner2004quantum}%
  \BibitemOpen
  \bibfield  {author} {\bibinfo {author} {\bibfnamefont {Crispin}\ \bibnamefont
  {Gardiner}}\ and\ \bibinfo {author} {\bibfnamefont {Peter}\ \bibnamefont
  {Zoller}},\ }\href@noop {} {\emph {\bibinfo {title} {Quantum noise: a
  handbook of Markovian and non-Markovian quantum stochastic methods with
  applications to quantum optics}}},\ Vol.~\bibinfo {volume} {56}\ (\bibinfo
  {publisher} {Springer Science \& Business Media},\ \bibinfo {year}
  {2004})\BibitemShut {NoStop}%
\bibitem [{\citenamefont {Breuer}\ and\ \citenamefont
  {Petruccione}(2002)}]{breuer2002theory}%
  \BibitemOpen
  \bibfield  {author} {\bibinfo {author} {\bibfnamefont {Heinz-Peter}\
  \bibnamefont {Breuer}}\ and\ \bibinfo {author} {\bibfnamefont {Francesco}\
  \bibnamefont {Petruccione}},\ }\href@noop {} {\emph {\bibinfo {title} {The
  theory of open quantum systems}}}\ (\bibinfo  {publisher} {Oxford University
  Press on Demand},\ \bibinfo {year} {2002})\BibitemShut {NoStop}%
\bibitem [{\citenamefont {Haroche}\ and\ \citenamefont
  {Raimond}(2006)}]{haroche2006exploring}%
  \BibitemOpen
  \bibfield  {author} {\bibinfo {author} {\bibfnamefont {Serge}\ \bibnamefont
  {Haroche}}\ and\ \bibinfo {author} {\bibfnamefont {Jean-Michel}\ \bibnamefont
  {Raimond}},\ }\href@noop {} {\emph {\bibinfo {title} {Exploring the quantum:
  atoms, cavities, and photons}}}\ (\bibinfo  {publisher} {Oxford university
  press},\ \bibinfo {year} {2006})\BibitemShut {NoStop}%
\end{thebibliography}%

\end{document}